# Blue and red shifted temperature dependence of implicit phonon shifts in Graphene

Sarita Mann and V.K. Jindal[1]

Department of Physics, Panjab University Chandigarh-160014, India

We have calculated the implicit shift for various modes of frequency in a pure graphene sheet. Thermal expansion and Grüneisen parameter which are required for implicit shift calculation have already been studied and reported. For this calculation, phonon frequencies are obtained using force constants derived from dynamical matrix calculated using VASP code where the density functional perturbation theory (DFPT) is used in interface with phonopy software. The implicit phonon shift shows an unusual behavior as compared to the bulk materials. The frequency shift is large negative (red shift) for ZA and ZO modes and the value of negative shift increases with increase in temperature. On the other hand, blue shift arises for all other longitudinal and transverse modes with a similar trend of increase with increase in temperature. The **q** dependence of phonon shifts has also been studied. Such simultaneous red and blue shifts transverse or out plane modes and surface modes, respectively leads to speculation of surface softening in out of plane direction in preference to surface melting.

## 1. INTRODUCTION

Graphene is a 2D thermodynamically stable structure of carbon. Graphene is of great importance due to its unique electronic and thermodynamic properties. It is the lightest and strongest material [1] with ability to efficiently conduct heat and electricity. Its unique electronic band structure is responsible for high intrinsic mobility and thus electron transport is governed by Dirac-like equation [2]. Graphene, a zero band gap semiconductor can be doped with B/N to modify its band gap and optical behavior [3, 4] for integrated optical and logical circuits. After its first synthesis [5], enormous experimental activity also supplemented theoretical results on many of the electronic, optical, thermal and thermodynamical properties.

Graphene is unusual material because it has a negative overall thermal expansion [6-15]. A lot of theoretical and experimental research is ongoing in this area. In our previous work, we have studied various thermodynamic properties like specific heat, entropy and free energy in harmonic approximation and also presented results on the phenomena of thermal expansion [15]. The negative thermal expansion in graphene is due to highly negative value of Grüneisen parameter for ZA (out of plane transverse acoustic) mode near Γ point [16].

---

[1] Email for correspondence: Jindal@pu.ac.in

Negative thermal expansion has consequences in dictating implicit phonon frequency shifts with temperature. The phonons are affected by temperature primarily due to anharmonicity of the interaction potential. The overall result appears as a shift in the value of the phonon frequency as well as a broadening of the phonon profile usually termed as phonon width. Anharmonicity affects phonons in two ways to the shifts- firstly and dominantly through thermal expansion and secondly (generally oppositely) through explicit increased vibrational amplitudes of atoms governed by anharmonicity. The first dominant term is also usually called as implicit shift and the less dominant contribution arising out of explicit anharmonic terms in the potential energy as explicit shift. The calculation of explicit shift is huge computational task requiring multiple Brillouin Zone summations over complicated anharmonic terms. At temperatures significantly lower than the melting temperature, such contributions are at most $1/5^{th}$ of the implicit shift as observed in the case of naphthalene [17]. We deal here with the dominant implicit shifts that result from thermal expansion and Grüneisen parameter.

For graphene, which has been found to be a material with negative as well as positive Grüneisen parameters and negative thermal expansion, the implicit shift in phonon frequencies could be decided by the sign of Grüneisen parameter. It will be positive instead of negative for all modes which have positive Grüneisen parameters (in contradiction to most 3D bulk materials) and negative for those modes having negative Grüneisen parameters. Thus instead of a usual negative value, the implicit shift in graphene can be both, a large positive value (blue shift) and a large negative value (red shift), depending upon the negative or positive values of thermal expansion and Grüneisen parameter. Here, the Grüneisen parameters are negative for flexural modes and positive for surface modes. This causes unusual critical mode dependent implicit shift. The aim of this paper is to focus on implicit shift of various modes which may have positive or negative Grüneisen parameters.

Although thermal expansion, thermal conductivity and various thermodynamic properties have been studied well, the temperature dependence of phonon frequencies has not been extensively studied. The experimental studies pertaining to the temperature dependence of the phonon frequencies have been made for G-mode at Γ point and have been reported earlier [8-11]. Calizo et. al. [8] have studied the temperature dependence of G-mode frequency for single layer and bilayer graphene and found it to decrease with increase in temperature. Yoon et al. [9] had found the Raman shift in G-mode by eliminating the substrate effect and found it to be negative in the low temperature region. Pan *et. al.* [10] has shown a positive shift in the G-mode position in the temperature range of 300-1200K of the graphene samples prepared on a BN substrate. Linas et al.[11] have also carried out Raman scattering experiments and studied the shift in G-mode of graphene in the temperature range of 150K -800K and analyzed the G-band measurements by employing corrections using various theoretical models.

Theoretically, Bonini *et. al.* [12] has predicted a red shift in the G-mode frequency using DFT implemented in quantum espresso implementation. Anees *et. al.* [13] has presented temperature dependent study of phonons using molecular dynamics (MD) and quasiharmonic approximation (QHA). They report that LO/TO mode are blue shifted and ZO mode is red shifted in QHA while blue shift of ZO mode is reported in MD simulation. We also noted another comprehensive and interesting calculation of phonon dispersion for all the six modes at various temperatures by Koukaras et al [14] using molecular dynamics to simulate the trajectories and velocities of the atoms. The dispersion curves are obtained from the maxima in power spectral density. Their method is based on assuming that anharmonic effects and so also temperature effects are embedded in these through velocities etc. and thus they present temperature dependence at all points of the phonon curves of all branches. This is clearly an alternate approach to a typical anharmonic self energy approach where different anharmonic terms are included depending upon the temperature in question. They observed a red or blue shift in phonon frequencies with temperature depending on the choice of potential.

We have made an attempt to calculate the implicit shift in phonon frequencies of every phonon branch at M and K points and for LO/TO and ZO modes at Γ point for pure graphene. Our approach is simple and direct, based on analytical expressions, once the essential ingredients have been calculated. In this approach anharmonic terms to the lowest order, in particular cubic anharmonicity is retained that takes into account dominant implicit anharmonicity. The approach of [14] differs from this and thus it will be useful to present the results of the temperature dependence based on the two approaches to have a better insight. The paper is divided into four sections as follows: The introduction is followed by computational procedure for calculation of implicit shift along with theory behind the process details of VASP software used in calculations. Section 3 includes the results of Grüneisen parameters and volume expansion that are key ingredients for implicit shift calculation, along with results of implicit shift for various modes of vibration which are followed by conclusions in section 4.

## 2. COMPUTATIONAL PROCEDURE

### 2.1) Theory – basic expressions

As discussed in earlier section, implicit shift can be obtained from volume dependence of the phonons. The first and foremost step for implicit shift calculation is to obtain phonon frequencies. The phonon frequencies calculation for pure graphene has been done earlier [15] using VASP (Vienna *ab- initio* Simulation Package) [18-20] code which is a density functional theory (DFT) based code in combination with phonopy [21] code. We had shown that our phonon dispersion curve matches well with the experimental studies and theoretical estimates. Under QHA, the implicit shift in the phonon frequencies result as a consequence of thermal expansion which is further governed by the phonon frequencies and their dependence on volume. The implicit shift of phonon frequencies is defined [17] as

$$\left(\frac{\partial \omega_{\mathbf{q}j}}{\partial T}\right)_{im} = \frac{\partial \omega_{\mathbf{q}j}}{\partial V}\frac{\partial V}{\partial T} \tag{1}$$

Grüneisen parameter is calculated as $\quad \gamma_{\mathbf{q}j} = -\frac{V_0}{\omega_{\mathbf{q}j}}\frac{\partial \omega_{\mathbf{q}j}}{\partial V} \tag{2}$

Where $V_0$ is equilibrium volume and $\omega_{\mathbf{q}j}$'s the harmonic phonon frequencies at phonon wave vector $\mathbf{q}$ for the jth mode of vibration.

Thus implicit shift in terms of Grüneisen parameter is expressed [17] as

$$\left(\frac{\partial \omega_{\mathbf{q}j}}{\partial T}\right)_{im} = -\gamma_{\mathbf{q}j}\omega_{\mathbf{q}j}\beta \tag{3}$$

Where $\beta$ is the volume thermal expansion coefficient, given by

$$\beta = \frac{1}{V_0}\frac{\partial V}{\partial T} \tag{4}$$

Assuming that Grüneisen parameters are independent of temperature, eq. (3) has been integrated [17] to get the implicit frequency shift expression in phonons given by

$$\left(\frac{\Delta \omega_{\mathbf{q}j}}{\omega_{\mathbf{q}j}}\right)_{im} = \exp[-\gamma_{\mathbf{q}j}\Delta\varepsilon(T)] - 1 \tag{5}$$

Where $\varepsilon(T) = \int_0^T \beta dT$ is the volume thermal expansion and $\Delta\varepsilon(T)$ is the difference in thermal expansion at any temperature from its value at T=0K.

Thus two ingredients are required in the calculation of implicit shift in phonons, mode dependent Grüneisen parameters and volume thermal expansion. The mode dependent Grüneisen parameters (given by equation 2) provide a clear understanding of thermal expansion and various thermodynamic properties of graphene. The thermal expansion behavior of graphene is unusual with negative thermal expansion coefficient in low temperature region. This behavior is mostly governed by large negative Grüneisen parameters of ZA mode at all q-points which has been studied earlier [22].

The expression for quasi harmonic free energy [17] is given by

$$F_{qh} = \varphi(V) + k_B T \sum_{\mathbf{q},j} \ln\left[2\sinh\left(\frac{\hbar\omega_{\mathbf{q}j}}{2k_B T}\right)\right] \tag{6}$$

Where $\varphi(V)$ is static contribution of lattice to free energy and $\omega_{\mathbf{q}j}$'s the harmonic frequencies. The strained volume can be expressed in terms of unstrained volume $V_0$ as $V_0(1+\varepsilon)$,

Thus $\varphi(V)$ can be written as

$$\varphi(V) = \varphi(V_0) + \frac{1}{2}\varepsilon^2 V_0^2 \left(\frac{\partial^2 \varphi}{\partial V^2}\right) = \varphi(V_0) + \frac{1}{2}\varepsilon^2 V_0 B \tag{7}$$

Where $B = V_0\left(\frac{\partial^2 \varphi}{\partial V^2}\right)$ is the bulk modulus.

The condition of minimum energy $\frac{\partial F_{qh}}{\partial \varepsilon} = 0$, gives the expression for volume thermal expansion [17] as

$$\varepsilon = \frac{1}{2V_0 B}\sum_{\mathbf{q}j}\gamma_{\mathbf{q}j}\hbar\omega_{\mathbf{q}j}\coth\frac{\hbar\omega_{\mathbf{q}j}}{2kT} \tag{8}$$

**2.2) Computational details**

The ab-initio calculations are done using VASP software to find dynamical matrix and hence force constant calculations are made. It is a density functional theory [23] based software where projector augmented wave pseudo potentials (PAW) [24, 25] are employed in calculations. The PBE (Perdew-Burke-Ernzerhof) [26] exchange correlation (XC) functional of GGA (generalized gradient approximation) is adopted and the plane wave cut-off energy of 750eV is used in the calculations. The 4x4 supercell of graphene (32 atoms sheet) has been used for simulation and an interlayer separation of 12Å distance between two graphene sheets along perpendicular direction is used to avoid interlayer interactions. The Monkhorst-pack scheme is used for K-space sampling. Firstly the structure is relaxed under the condition that Hellmann-Feynman forces should be less than 0.005 eV/Å with a Γ centered (7 x 7 x 1) k-mesh. The tetrahedron method with Blöchl corrections [24] is used for treating the partial occupancies. The dynamical matrix obtained using VASP software is used in combination with phonopy software to obtain phonon frequencies. The force constants (FCs) are obtained by making use of density functional perturbation theory (DFPT). The bulk modulus is obtained by calculating second derivative of energy volume curve. Using equation 2 and 8, mode dependent Grüneisen parameters and volume thermal expansion are calculated respectively using FORTRAN programming. Finally the implicit shifts in phonon frequencies are calculated using equation 5.

## 3. RESULTS AND DISCUSSION

The phonon frequencies and hence phonon dispersion curve and various thermodynamic properties of pure graphene have already been studied. The phonon frequencies at high symmetry points of the Brillion zone calculated using different approximations (local density approximation (LDA), GGA-91 and GGA-PBE) implemented using VASP software are given in table 1 of our earlier work [15]. The quasiharmonic study for the free energy minimization and further calculations have been made here to compute implicit shift in the phonon frequencies by making use of GGA-PBE potential.

**3.1) Mode dependent Grüneisen parameters**

Although study of mode dependent Grüneisen parameters using equation 2 have already been done [22], but the results are also reproduced here for the sake of clarity as they enter into implicit frequency shift calculation directly. In an earlier study, attempt has been made to figure out the negative Grüneisen parameters of ZA mode which greatly contribute to negative thermal expansion in pure graphene at low temperature since optical modes are not excited at low temperatures.

It is clear from figure 1 that ZA and ZO modes have negative Grüneisen parameters although the value of Grüneisen parameter is large and negative for ZA mode near Γ point while its value for ZO mode is close to zero. At M and K points, their Grüneisen parameters are nearly same having small negative value. All other phonon frequency modes have small but positive Grüneisen parameters with an average value close to 1. These values have direct impact on implicit shift in phonon frequencies which are further calculated using equation 5. The branch dependent Grüneisen parameters obtained here matches well with a recent study on lattice thermal transport properties [16].

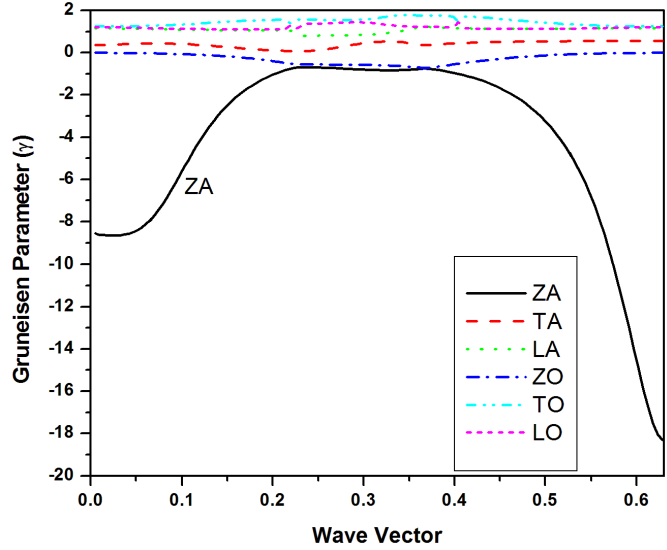

Figure 1. Branch dependent Grüneisen parameters for pure graphene [22]

### 3.2) Volume Thermal Expansion

The volume of graphene sheet itself has no meaning since the volume is arbitrarily taken as area of graphene sheet multiplied by interlayer separation. Since the volume of unit cell has a constant interlayer separation, the change in volume to original volume is same as change in surface area of sheet to the original surface area. The volume expansion calculated using equation 8 is a function of temperature. Thus we have plotted the net change in volume expansion at any temperature as net expansion w.r.t. its value at 0K and the corresponding curve is shown in figure 2. The volume expansion shows that there is a continuous decrease in volume of the unit cell with increase in temperature. This indicates that contraction of graphene occurs with increase in temperature. It may be pointed out that this contraction in volume is the result of lowest order calculation. As soon as higher anharmonic terms are included, a positive contribution will keep on building up increasing with temperature and positive expansion will be the net result at high enough temperatures.

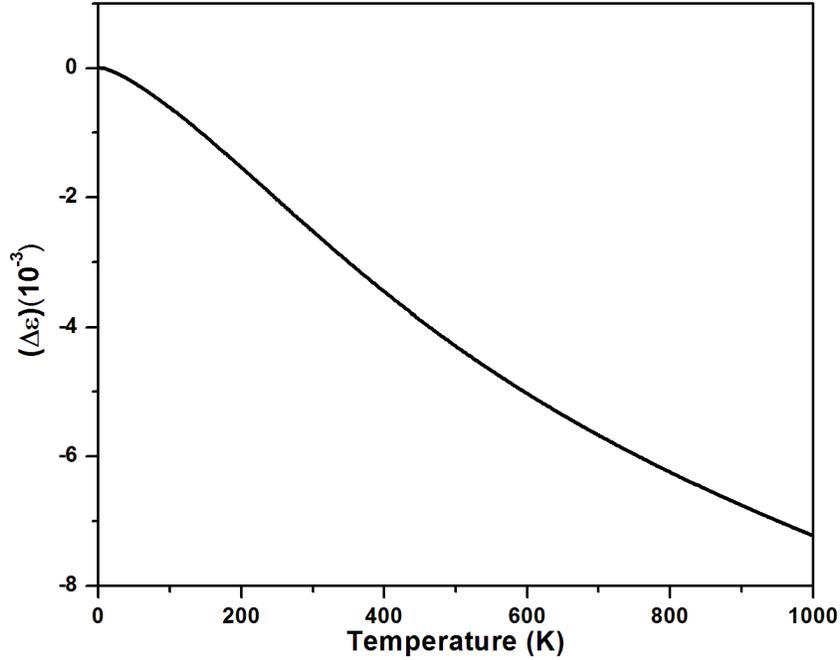

Figure 2. Change in volume thermal expansion $\Delta\varepsilon(T)$ as a function of temperature for pure graphene

**3.3) Implicit Shift**

It is now trivial to calculate implicit frequency shift using equation 5, where the essential ingredients which enter into expression have already been calculated. The Grüneisen parameters are assumed to be constant for each branch (temperature independent). Thus implicit shift is calculated individually for each phonon branch at M, K and Γ points of the Brillouin zone separately. Figure 3(a) shows the implicit shift for all branches at M point while figure 3(b) shows the shift at K point. Since acoustic modes (LA, TA and ZA) vanish at Γ point, therefore, we only show the shift for optic modes: (LO/TO) and ZO modes at Γ point in figure 4.

Since the Grüneisen parameter is negative for ZA and ZO modes as can be seen in figure 1, the corresponding shift comes out to be negative as it involves the product of Grüneisen parameter and thermal expansion (equation 5) and its negative value increases with increase in temperature. Further ZA mode Grüneisen parameter strongly depends on phonon wave vector **q** (figure 1). The Grüneisen parameters for ZA and ZO modes are nearly the same between the **q** values ranging from 0.233 Å$^{-1}$ to 0.467 Å$^{-1}$. This range of **q** values includes M and K points of Brillouin zone. Therefore there is no significant difference between the shifts of ZA and ZO modes at M and K points. The Grüneisen parameter of ZA mode has very large effect near Γ point but we cannot present the shift in ZA mode at Γ point because acoustic mode frequency itself vanishes.

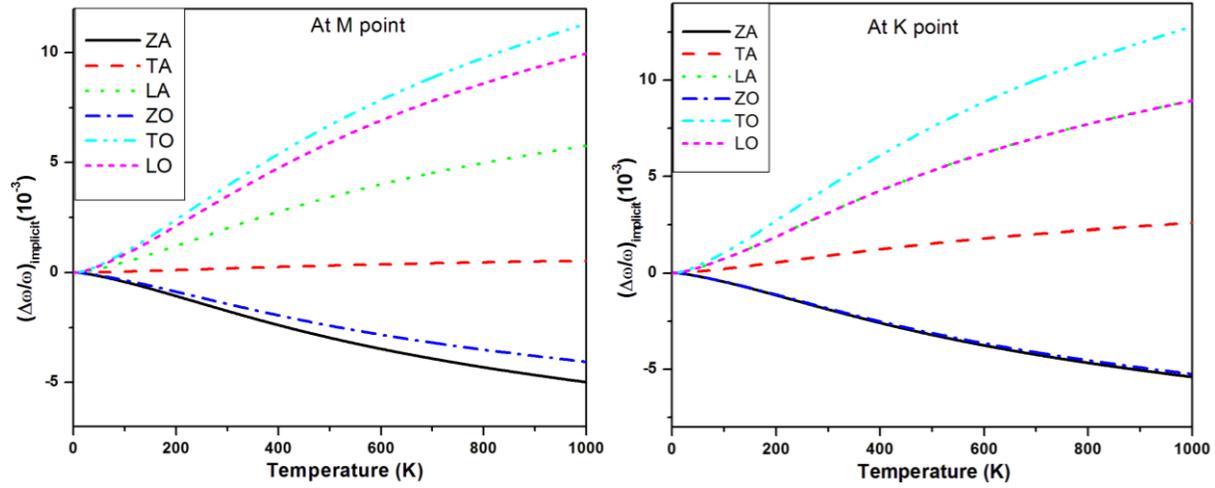

Figure 3. Implicit shift in phonon frequency for all branches (a) at M point (b) at K point

The implicit shift of all other longitudinal and transverse modes (surface modes) is positive and increases with increase in temperature. This is because these modes have positive value of Grüneisen parameters. Among all in-plane modes, the frequency shift is lowest for TA mode and highest for LO and TO modes. Thus the behavior of ZA and ZO modes (out of surface modes) is entirely opposite to the behavior of longitudinal and transverse acoustic and optical branches originating from surface modes.

The LO and TO modes are degenerate at $\Gamma$ point and they correspond to G-mode frequency which is Raman active in graphene. The implicit shifts for LO/TO and ZO modes are also calculated at $\Gamma$ point and results are compared with an experimental study of G-mode shift and a theoretical study, which is shown in figure 4.

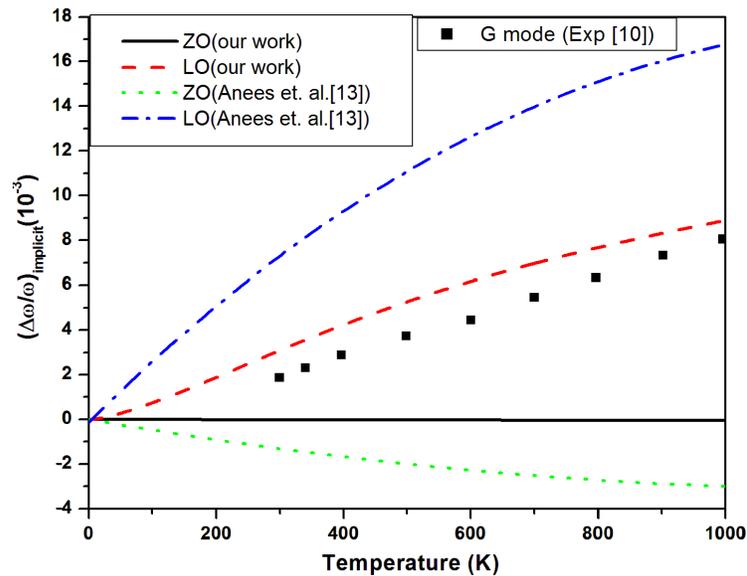

Figure 4. Implicit shift in phonon frequency for LO/TO mode and ZO mode at Γ point compared with experimental data of G-mode shift [10] and theoretical estimate [13].

We have found a blue shift in LO/TO frequency in the temperature range of 0-1000K. Since Grüneisen parameter for LO/TO mode is positive, the implicit shift are also positive for these modes and this is unusual behavior when compared to normal 3D materials. In most of the experimental work [8-11], Raman shift of G-mode has been studied. Earlier studies [8-9] were restricted to a narrow temperature range of 0-400K and measured a red shift in G-mode. However recently, Pan et. al. [10] have studied the temperature dependent shift of G mode in graphene/BN hetero-structures in a broad range of temperature (200-1300K) and measured a blue shift in frequency in the whole temperature range. Linas et. al. [11] has also measured a red shift in G-mode in the temperature range of 150-800K. They corrected for the substrate to predict blue shift. Recent theoretical works [13,14] report contradictory results. Whereas [13] predicted a blue shift in the LO/TO mode (G-mode) even at low temperatures, [14] reported red shift using Lindsay-Broidoy-Tersoff-2010 potential and blue shift at low temperatures using LCBOP and AIREBO potentials (fig 3 in their paper). It would be interesting to have a look at the results of thermal expansion from their averaged behavior of position vector in [14] whether this is positive or negative for temperatures well below 1000K. The two reported studies [10] and [13] show a blue shift in LO/TO mode frequency which are compared with our data. The comparison of LO/TO mode frequency in figure 4 shows that the shift predicted by our method is close to experimental study in the entire range of temperature, while it is underestimated as compared with theoretical data by Anees et al. [13]. The data presented here from reference [10] and [13] has been read from the appropriate shifts presented and shown as relative shift here by taking their phonon frequencies at T=0K. All our shifts presented here are relative shifts.

We would like to focus primarily on three contributions addressing this issue [10], for experimental measurements, [13] for theoretical calculations, and [16] for a recent study on Grüneisen parameters. In view of wide range in contradictions appearing in literature regarding shift measurements, it is important to conjecture the merits of these three contributions alongwith the present work.

For the ZO mode, Anees et. al. [13] predicted a red shift for these in quasi harmonic approximation ZO. We have also obtained a red shift in the frequency for ZO mode at Γ point and its negative value increases with increase in temperature. Our work is a direct consequence of Grüneisen parameters of pure graphene. Since Grüneisen parameters are negative for ZO mode and are expected to be a constant quantity w.r.t. temperature, their behavior is different from normal modes for any material. The implicit shift for this mode is negative but the value of shift for ZO mode is quite low as compared to shifts in LO/TO modes. There is no reported measurement for ZO mode shift in literature as ZO mode is Raman and IR inactive. On comparison of numerical values, our results are significantly lower. Although the ZO mode frequency predicted by them [13] at 0K is 0.4% lower as compared to experimental value [27] while our frequency closely matches with experiment. If their relative shift is calculated using experimental frequencies, their results of the relative shift for ZO modes tend to agree with ours. It has been observed that ZO mode shift predicted by Anees et. al. [13] closely matches with our ZO mode shifts at M and K points in contrast to Koukaras et al [14]. But the Grüneisen parameters for ZO mode at Γ point are having much lower magnitude compared to their value at M and K points. Our value of Grüneisen parameters for ZO mode at Γ point matches well with a recent study [16].Thus the ZO mode shift should be appreciable at M and K points (figure 3) and not at the Γ point Thus the value of Grüneisen parameters at different **q** points confirms that the shift for ZO mode should be appreciable at the regions between M and K points only.

## 4. SUMMARY AND CONCLUSIONS

We have studied the implicit shift in phonon frequency for various modes of vibration for pure graphene at Γ, M and K points. There is a negative implicit shift in ZA and ZO modes of vibration whose magnitude increases with increase in temperature for sufficiently high temperatures above room temperature, while other modes of vibration show unusual positive implicit shift which also increases with increase in temperature. We have compared our data with available experimental and theoretical results at Γ point. Our results find good agreement for available measured shift. We have been able to segregate the relevant and more likely results from other authors on the basis of established behavior of Grüneisen parameters and thermal expansion.

Our results for the relative implicit shift indicate a behavior which is opposite to usually observed behavior in 3D materials. Around temperatures upto 1000K, the contribution from explicit anharmonic terms is not expected to be significant as the melting temperature of graphene is very high. The results in shift of phonons modes in graphene become interesting on two counts. Firstly, in the same material for some modes, the shifts are red and for other, these are blue shifted. The blue shift is unusual behavior which seems to be a general behavior of 2D materials having negative thermal expansion. This is so for surface modes and leads to hardening rather than softening of the modes. On the other hand, the out of plane modes soften on increase in temperature. It probably leads to two kinds of melting of such materials, surface melting and out of plane melting dictated differently.

## 5. ACKNOWLEDGEMENTS

The authors of this paper are highly grateful to VASP team and the phonopy team for providing the code, the HPC facilities at IUAC (New Delhi) and the computing facilities at Department of Physics, Panjab

University, Chandigarh. VKJ also acknowledges academic support from Guru Jambheshwar University for the Honorary Professor position.University, Chandigarh. VKJ also acknowledges academic support from Guru Jambheshwar University for the Honorary Professor position.